%% file: main.tex
\documentclass[acmsmall]{acmart}

\usepackage{booktabs}
\usepackage{footnote}

\usepackage{amsmath}
\usepackage{mathrsfs}

\usepackage{epstopdf}
\usepackage{multirow}
\usepackage[skip=0pt]{subcaption}
\usepackage{soul}

\usepackage{tabularx}
 
\usepackage{enumitem}

\renewcommand\vec[1]{\overrightarrow{#1}}

\usepackage{microtype}
\usepackage[switch]{lineno}

\usepackage{hyperref}

\AtBeginDocument{%
  \providecommand\BibTeX{{%
    \normalfont B\kern-0.5em{\scshape i\kern-0.25em b}\kern-0.8em\TeX}}}

\setcopyright{acmcopyright}
\copyrightyear{2018}
\acmYear{2018}
\acmDOI{10.1145/1122445.1122456}

\acmConference[ACM TKDD]{ACM Transactions on Knowledge Discovery from Data}{New York}{NY, USA, 2021}
\acmBooktitle{ACM Transactions on Knowledge Discovery from Data}
\acmPrice{15.00}
\acmISBN{978-1-4503-XXXX-X/18/06}

\begin{document}

\title[Corpus-level and Concept-based Explanation]{Corpus-level and Concept-based Explanations for Interpretable Document Classification}

\author{Tian Shi}
\email{tshi@vt.edu}
\affiliation{
  \institution{Virginia Tech}
  \country{USA}
}
\author{Xuchao Zhang}
\email{xuczhang@vt.edu}
\affiliation{
  \institution{Virginia Tech}
  \country{USA}
}
\author{Ping Wang}
\email{ping@vt.edu}
\affiliation{
  \institution{Virginia Tech}
  \country{USA}
}
\author{Chandan K. Reddy}
\email{reddy@cs.vt.edu}
\affiliation{
  \institution{Virginia Tech}
  \country{USA}
}

\renewcommand{\shortauthors}{Trovato and Tobin, et al.}

\begin{abstract}
 Using attention weights to identify information that is important for models' decision making is a popular approach to interpret attention-based neural networks. This is commonly realized in practice through the generation of a heat-map for each single document based on attention weights. However, this interpretation method is fragile and easy to find contradictory examples. In this paper, we propose a corpus-level explanation approach, which aims to capture causal relationships between keywords and model predictions via learning importance of keywords for predicted labels across a training corpus based on attention weights.
 Based on this idea, we further propose a concept-based explanation method that can automatically learn higher level concepts and their importance to model prediction task. Our concept-based explanation method is built upon a novel Abstraction-Aggregation Network, which can automatically cluster important keywords during an end-to-end training process. We apply these methods to the document classification task and show that they are powerful in extracting semantically meaningful keywords and concepts.
Our consistency analysis results based on an attention-based Na\"ive Bayes classifier also demonstrate these keywords and concepts are important for model predictions.
\end{abstract}



\maketitle

\input{introduction}
\input{relatedwork}
\input{model}

\input{experiments}

\input{conclusion}

\begin{acks}
This work was supported in part by the US National Science Foundation grants  IIS-1707498, IIS-1838730, and NVIDIA Corporation.
\end{acks}

\bibliographystyle{ACM-Reference-Format}
\bibliography{ref}

\appendix

\end{document}

%% file: introduction.tex
\section{Introduction}
\label{sec:intro}

Attention Mechanisms \cite{bahdanau2014neural} have boosted performance of deep learning models in a variety of natural language processing (NLP) tasks, such as sentiment analysis \cite{wang2016attention,pontiki2016semeval}, semantic parsing \cite{vinyals2015grammar}, machine translation \cite{luong2015effective}, reading comprehension \cite{hermann2015teaching,devlin2019bert} and others.
Attention-based deep learning models have been widely investigated not only because they achieve state-of-the-art performance, but also because they can be interpreted by identifying important input information via visualizing heat-maps of attention weights \cite{strobelt2018seq2seq,vig2019visualizing,ghaeini2018interpreting}, namely attention visualization.
Therefore, attention mechanisms help end-users to understand models and diagnose trustworthiness of their decision making.

However, the attention visualization approach still suffers from several drawbacks:
1) The fragility of attention weights can easily make end-users find contradicting examples, especially for noisy data and cross-domain applications. For example, a model may attend on punctuation or stop-words.
2) Attention visualization cannot automatically extract high-level concepts that are important for model predictions.
For example, when a model assigns news articles to \textit{Sports}, relevant keywords may be \textit{player}, \textit{basketball}, \textit{coach}, \textit{nhl}, \textit{golf}, and \textit{nba}.
Obviously, we can build three concepts/clusters for this example, i.e., roles (\textit{player}, \textit{coach}), games (\textit{basketball}, \textit{soccer}), and leagues (\textit{nba}, \textit{nhl}).
3) Attention visualization still relies on human experts to decide if keywords attended by models are important to model predictions.

There have been some studies that attempt to solve these problems.
For example, \citet{jain2019attention,serrano2019attention} focused on studying if attention can be used to interpret a model, however, there are still problems in their experimental designs \cite{wiegreffe2019attention}.
\citet{yeh2019concept} tried to apply a generic concept-based explanation method to interpret BERT models in the text classification task, however, they did not obtain semantically meaningful concepts for model predictions.
\citet{antognini2021rationalization} introduced a concept explanation method that first extracts a set of text snippets as concepts and infers which ones are described in the document, and then it explained the
predictions of sentiment with a linear aggregation of concepts.
In this paper, we propose a general-purpose corpus-level explanation method and a concept-based explanation method based on a novel Abstraction-Aggregation Network (AAN) to tackle the aforementioned drawbacks of attention visualization.
We summarize primary contributions of this paper as follows:
\begin{itemize}[leftmargin=*,topsep=0pt,itemsep=1pt,partopsep=1pt, parsep=1pt]

\item To solve the first problem, we propose a \textit{corpus-level explanation} method, which aims to discover causal relationships between keywords and model predictions. 
The importance of keywords is learned across the training corpus based on attention weights.
Thus, it can provide more robust explanation compared to attention visualization case studies.
The discovered keywords are semantically meaningful for model predictions.

\item To solve the second problem, 
we propose a \textit{concept-based explanation method} (case-level and corpus-level) that can automatically learn semantically meaningful concepts and their importance to model predictions.
The concept-based explanation method is based on an AAN that can automatically cluster keywords, which are important to model predictions, during the end-to-end training for the main task.
Compared to the basic attention mechanisms, the models with AAN do not compromise on classification performance or introduce any significant number of new parameters.

\item To solve the third problem, we build a \textit{Na\"ive Bayes Classifier} (NBC), which is based on an \textit{attention-based bag-of-words document representation} technique and the causal relationships discovered by the corpus-level explanation method.
By matching predictions from the model and NBC, i.e., consistency analysis, we can verify if the discovered keywords are important to model predictions.
This provides an automatic verification pipeline for the results from the corpus-level explanation and concept-based explanation methods.

\end{itemize}

The rest of this paper is organized as follows:
In Section \ref{sec:relate_work}, we introduce related work of feature-based and concept-based explanation.
In Section \ref{sec:methods}, we first present details of our proposed abstraction-aggregation network (AAN), and then discuss corpus-level and concept-based explanation methods.
In Section \ref{sec:exp}, we evaluate different self-attention and AAN based models on three different datasets.
We also show how corpus-level and concept-based explanations can help us interpret attention-based classification models and understand training corpus.
Our discussion concludes in Section~\ref{sec:conclusion}.

%% file: relatedwork.tex
\section{Related Work}
\label{sec:relate_work}

Increasing interpretability on machine learning models has become an important topic of research in recent years. Most prior work \cite{gilpin2018explaining, liu2019incorporating, lundberg2017unified} focus on interpreting models via feature-based explanations, which alters individual features such as pixels and word-vectors in the form of either deletion \cite{ribeiro2016model} or perturbation \cite{sundararajan2017axiomatic}. However, these methods usually suffer from the reliability issues when adversarial perturbations \cite{ghorbani2019interpretation} or even simple shifts \cite{kindermans2019reliability} in the input.
Moreover, the feature-based approaches explain the model behavior locally \cite{ribeiro2016model} for each data samples without a global explanation \cite{kim2018interpretability,ghorbani2019towards} on how the models make their decisions.
In addition, feature-based explanation is not necessarily the most effective way for human understanding.

To alleviate the issues of feature-based explanation models, some researches have focused on explaining the model results in the form of high-level human concepts \cite{zhou2018interpretable,tjoa2020survey,das2020opportunities,chen2020concept,wu2020towards,bodria2021benchmarking,zaeem2021cause}.
Unlike assigning the importance scores to individual features, the concept-based methods use the corpus-level concepts as the interpretable unit. For instance, concept ``wheels" can be used for detecting the vehicle images and concept ``Olympic Games" for identifying the sports documents. 
However, most of the existing concept-based approaches require human supervision in providing hand-labeled examples of concepts, which is labor intensive and some human bias can be introduced in the explanation process \cite{kim2018interpretability,tang2016tri,tang2019social}.
Recently, automated concept-based explanation methods \cite{yeh2019concept, bouchacourt2019educe} are proposed to identify higher-level concepts that are meaningful to humans automatically.
However, they have not shown semantically meaningful concepts on text data.
In text classification area, most of the existing approaches focus on improving the classification performance, but ignore the interpretability of the model behaviors \cite{yang2016hierarchical}. \citet{liu2019incorporating} utilize the feature attribution method to help users interpret the model behavior. \citet{bouchacourt2019educe} propose a self-interpretable model through unsupervised concept extraction. 
However, it requires another unsupervised model to extract concepts.
%
 
%
%
%
%
%
%
%
%
%
%
%
%
%
%

%% file: model.tex
\section{Proposed Work}
\label{sec:methods}

In this section, we first introduce the classification framework and our Abstraction-Aggregation Network (AAN).
Then, we systematically discuss the \textit{corpus-level explanation}, \textit{concept-based explanation}, and attention-based \textit{Na\"ive Bayes Classifier}.

\subsection{The Proposed Model}
\subsubsection{Basic Framework}
\label{sec:model_framework}

A typical document classification model is equipped with three components, i.e., an encoder, an attention or pooling layer and a classifier.
1)~\textbf{Encoder}: 
An encoder reads a document, denoted by $d=(w_1,w_2,...,w_T)$, and transforms it to a sequence of hidden states $H=(h_1,h_2,...,h_T)$.
Here, $w_t$ is the one-hot representation of token $t$ in the document.
$h_t$ is also known as a word-in-context representation.
Traditionally, the encoder consists of a word embedding layer followed by a LSTM \cite{hochreiter1997long} sequence encoder.
Recently, pre-trained language models \cite{devlin2019bert,yang2019xlnet,peters2018deep} have emerged as an important component for achieving superior performance on a variety of NLP tasks including text classification.
Our model is adaptable to any of these encoders.
2)~\textbf{Attention/Pooling}:
The attention or pooling (average- or max-pooling) layer is used to construct a high-level document representation, denoted by $v^\text{doc}$.
In attention networks, the attention weights show the contributions of words to the representations \cite{yang2016hierarchical,lin2017structured}.
Compared with pooling, attention operations can be well interpreted by visualizing attention weights \cite{yang2016hierarchical}.
3)~\textbf{Classifier}: 
The document representation is passed into a classifier to get the probability distribution over different class labels.
The classifier can be a multi-layer feed-forward network with activation layer followed by a softmax layer, i.e., $y=\text{softmax}(W_2\cdot\text{ReLU}(W_1\cdot v^\text{doc}+b_1)+b_2)$,
where $W_1, W_2, b_1$ and $b_2$ are model parameters.

\begin{figure}[!t]
	\centering
	\includegraphics[width=0.85\linewidth]{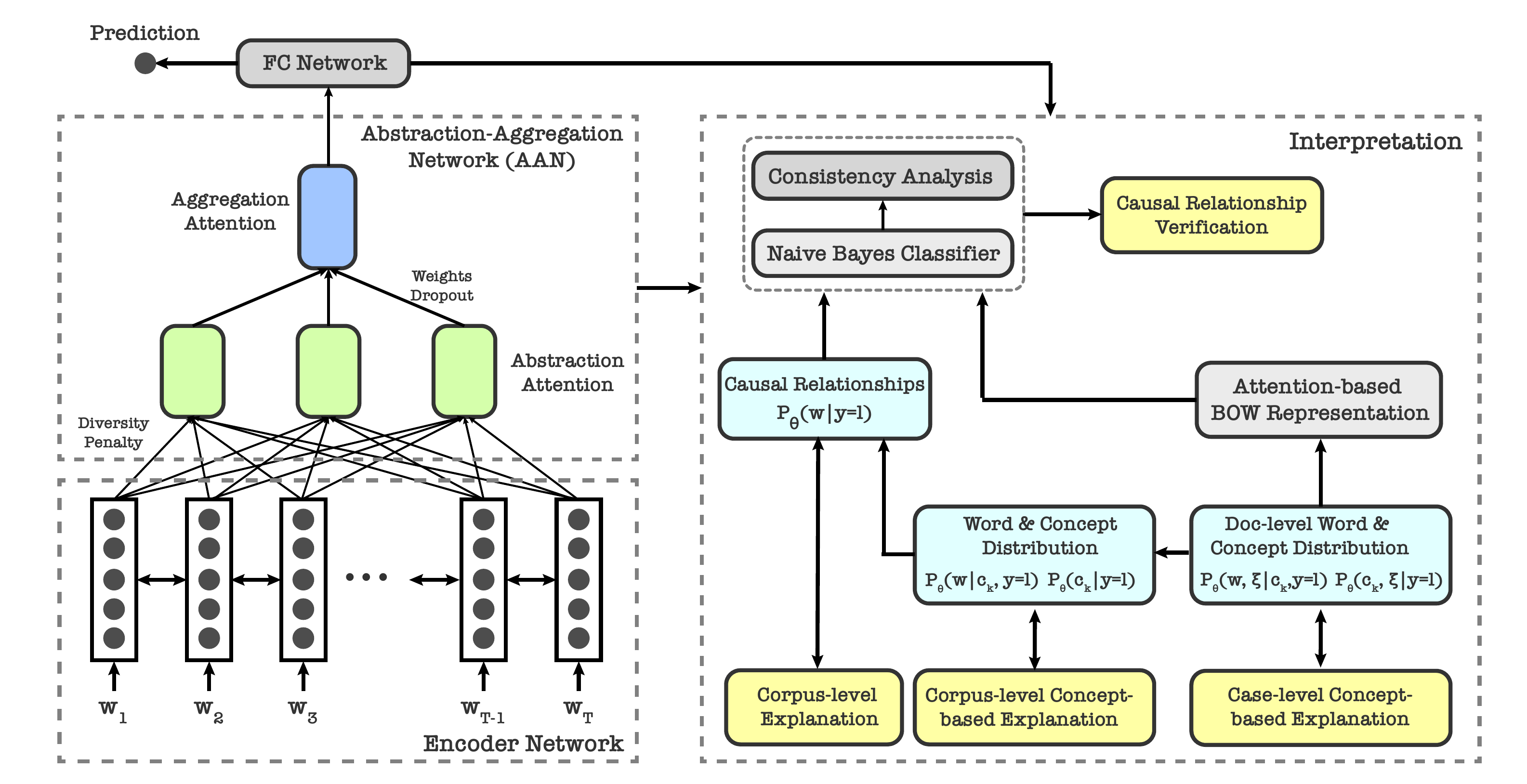}
	\caption{The proposed Abstraction-Aggregation Network and different interpretation methods.}
	\label{fig:model_framework}
	\vspace{-5mm}
\end{figure}
 
To infer parameters, we can minimize the averaged cross-entropy error between predicted and ground-truth labels. Here, loss function is defined as $\mathcal{L}_\theta=-\sum_{l=1}^L\hat{y}\log(y)$, where $\hat{y}$ represents the ground-truth label and $L$ is the number of class labels.
The model is trained in an end-to-end manner using back-propagation.


\subsubsection{Abstraction-Aggregation Network}

In order to use different explanation methods, especially concept-based explanation, to interpret deep neural networks,
we propose a novel AAN for the Attention/Pooling layer, which first captures keywords for different concepts from a document, and then aggregates all concepts to construct the document representation (see Fig.~\ref{fig:model_framework}).

AAN has two stacked attention layers, namely, \textit{abstraction-attention (abs)} and \textit{aggregation-attention (agg)} layers.
In the \textit{abs} layer, for each attention unit $k$, we calculate the alignment score $u^\text{abs}_{k,t}$ and attention weight $\alpha^\text{abs}_{k,t}$ as follows:
\begin{equation}
\aligned
u^\text{abs}_{k,t}&=(g^\text{abs}_k)^\top h_t,\\
\alpha^\text{abs}_{k,t}&=\frac{\exp(u^\text{abs}_{k,t})}{\sum_{\tau=1}^T\exp(u^\text{abs}_{k,\tau})},
\endaligned
\label{eqn:abstraction_attn}
\end{equation}
where $g^\text{abs}_k$ are model parameters.
Here, we do not apply linear transformation and tanh activation when calculating alignment scores for two reasons: 
1)~\textbf{Better intuition}: Calculating attention between $g^\text{abs}_k$ and $h_t$ in Eq.~(\ref{eqn:abstraction_attn}) is the same as calculating a normalized similarity between them.
Therefore, abstraction-attention can also be viewed as a clustering process, where $g^\text{abs}_k$ determines the centroid of each cluster.
In our model, concepts are related to the clusters discovered by AAN.
2)~\textbf{Fewer parameters}: Without linear transformation layer, abstraction-attention layer only introduces $K\times|h_t|$ new parameters, where $|h_t|$ is the dimension of $h_t$ and $K\ll|h_t|$.
The $k^\text{th}$ representation is obtained by
$v^\text{abs}_k=\sum_{t=1}^T \alpha^\text{abs}_{k,t} h_t$.
We use $K$ to denote the total number of attention units.

In the \textit{agg} layer, there is only one attention unit.
The alignment score $u^\text{agg}_k$ and attention weight $\alpha^\text{agg}_k$ are obtained by
$$
u^\text{agg}_k=(g^\text{agg})^\top\tanh(W_\text{agg} v^\text{abs}_k+b_\text{agg}),
$$
and
$$
\alpha^\text{agg}_k=\frac{\exp(u^\text{agg}_t)}{\sum_{\kappa=k}^K\exp(u^\text{agg}_\kappa)},
$$
where $W_\text{agg}, b_\text{agg}$ and $g^\text{agg}$ are model parameters.
The final document representation is obtained by
$v^\text{doc}=\sum_{k=1}^K \alpha^\text{agg}_k v^\text{abs}_k$.
It should be noted that AAN is different from hierarchical attention \cite{yang2016hierarchical}, which aims to get a better representation.
However, AAN is used to automatically capture concepts/clusters.
We have also applied two important techniques to obtain semantically meaningful concepts. \\ \vspace{-0.1in}

\noindent\textbf{1) Diversity penalty for abstraction-attention weights}: 
To encourage the diversity of concepts, we introduce a new penalization term to abstraction-attention weights $A=[\vec{\alpha}^\text{abs}_{1}, \vec{\alpha}^\text{abs}_{2},...,\vec{\alpha}^\text{abs}_{K}]\in\mathbb{R}^{T\times K}$, where $\vec{\alpha}^\text{abs}_{k}=(\alpha^\text{abs}_{k,1},\alpha^\text{abs}_{k,2},...,\alpha^\text{abs}_{k,T})^\top$.
We define the penalty function as
\begin{equation}
    \mathcal{L}_\text{div}=\frac{1}{K}\|A^\top A-I\|_F,
    \label{eqn:penalty}
\end{equation}
where $\|\cdot\|_F$ represents the Frobenius norm of a matrix.
Hence, the overall loss function is expressed as $\mathcal{L}=\mathcal{L}_\theta+\mathcal{L}_\text{div}$.\\ \vspace{-0.1in}

\noindent\textbf{2) Dropout of aggregation-attention weights}:
In the aggregation-attention layer, it is possible that $\alpha^\text{agg}_k\approx1$ for some $k$, and other attention weights tend to be $0$.
To alleviate this problem, we apply dropout with a small dropout rate to aggregation-attention weights $(\alpha^\text{agg}_1, \alpha^\text{agg}_2,...,\alpha^\text{agg}_K)$, namely attention weights dropout.
It should be noted that large dropout rate has negative impact to the explanation, since it discourages the diversity of concepts.
More specifically, the model will try to capture keywords in the dropped abstraction-attention units by the other units.

\subsection{Explanation}

In this section, we discuss corpus-level and concept-based explanations.
Given a corpus $\mathcal{C}$ with $|\mathcal{C}|$ documents, we use $d$ or $\xi$ to represent a document.
Let us also use $\theta$ to denote all parameters of a model and $\mathcal{V}$ to represent the vocabulary, where $|\mathcal{V}|$ is the size of $\mathcal{V}$.
Throughout this paper, we will assume that both prior document probability $p(d)$ and prior label probability $p_\theta(y=l)$ are constants.
For example, in a label-balanced dataset, $p_\theta(y=l)\approx 1/L$.

We will first apply the attention weights visualization technique to the proposed AAN model.
Here, the document representation can be directly expressed by the hidden states, i.e.,
\begin{equation*}
v^{\text{agg}}=\sum_{t=1}^T\left(\sum_{k=1}^K\alpha^\text{agg}_k\alpha^\text{abs}_{k,t}\right)h_t, 
\end{equation*}
where
\begin{equation}
    \alpha_t^d=\sum_{k=1}^K\alpha^\text{agg}_k\alpha^\text{abs}_{k,t}
    \label{eqn:attention_weight_aan}
\end{equation}
gives the contribution of word $w_t$ to the document representation.
Therefore, we can interpret single example via visualizing the combined weights $\alpha_t^d$.




\subsubsection{Corpus-Level Explanation}
\label{sec:corpus_level_explanation}

Corpus-level explanation aims to find casual relationships between keywords captured by attention mechanism and model predictions, which can provide robust explanation for the model.
To achieve this goal, we learn distributions of keywords for different predicted labels on training corpus based on attention weights.

Formally, for a given word $w\in\mathcal{V}$ and a label $l$ predicted by a model $\theta$\footnote{Here, the label is the model's prediction, not the ground-truth label, because our goal is to explain the model.}, the importance of the word to the label can be estimated by the probability
$p_\theta(w|y=l)$ across the training corpus $\mathcal{C}_\text{train}$ since the model is trained on it.
Therefore, $p_\theta(w|y=l)$ can be expanded as follows:
\begin{equation}
    p_\theta(w|y=l)= \sum_{\xi\in\mathcal{C}_\text{train}^l} p_\theta(w,\xi|y=l),
    \label{eqn:keywords_corpus_level}
\end{equation}
where $\mathcal{C}_\text{train}^l\subset\mathcal{C}_\text{train}$ consists of documents with model predicted label $l$.
For each document $\xi\in\mathcal{C}_\text{train}^l$, probability $p_\theta(w,\xi|y=l)$ represents the importance of word $w$ to label $l$, which can be defined using attention weights, i.e.,
\begin{equation}
    p_\theta(w,\xi|y=l):=\frac{\sum_{t=1}^T\alpha^\xi_t\cdot\delta(w_t,w)}{\sum_{\xi'\in\mathcal{C}_\text{train}}f_{\xi'}(w)+\gamma},
    \label{eqn:keywords_case_level}
\end{equation}
where $f_{\xi'}(w_t)$ is frequency of $w_t$ in document $\xi'$ and $\gamma$ is a smooth factor.
$\delta(w_t,w)=\begin{cases}1& \text{if } w_t=w\\ 0& \text{otherwise}\end{cases}$ is a delta function.
The denominator is applied to reduce noises from stop-words and punctuation.
For the sake of simplicity, we will use $p_\theta(w,l,\mathcal{C})$ to denote $p_\theta(w_t|y=l)$, where $\mathcal{C}$ corresponds to the corpus Eq.~(\ref{eqn:keywords_corpus_level}), and can be different from $\mathcal{C}_\text{train}$ in our applications.
The denominator in Eq.~(\ref{eqn:keywords_case_level}) is always determined by the training corpus.

As to applications: 1) Since Eq.~(\ref{eqn:keywords_corpus_level}) captures the importance of words to model predicted labels, we can use it as a criterion for finding their causal relationships.
In experiments, we can collect top-ranked keywords for each label $l$ for further analysis.
2) We can also use corpus-level explanation to measure the difference between two corpus (i.e., $\mathcal{C}_\text{test1}$ and $\mathcal{C}_\text{test2}$).
Formally, we can compare $\frac{|\mathcal{C}_\text{train}|}{|\mathcal{C}_\text{test1}|}\cdot p_\theta(w,l,\mathcal{C}_\text{test1})$ with $\frac{|\mathcal{C}_\text{train}|}{|\mathcal{C}_\text{test2}|}\cdot p_\theta(w,l,\mathcal{C}_\text{test2})$ across different words and class labels.
The difference can be evaluated by Kullback-Leibler divergence \cite{kullback1997information}.
In addition, we can get mutual keywords shared across different domains based on these distributions.

It should be noted that the corpus-level explanation discussed in this section can be applied to interpret different attention-based networks.

\subsubsection{Concept-Based Explanation}

The corpus-level explanation still suffers from the drawback that it cannot automatically obtain higher-level concepts/clusters for those important keywords.
To alleviate this problem, we propose concept-based explanation for our AAN model.
In AAN, each abstraction-attention unit can capture one concept/cluster.
Here, we will take distribution of concepts into consideration.
Formally, we express $p_\theta(w_t|y=l)$ as follows:

$$
p_\theta(w|y=l)=\sum_{k=1}^K p_\theta(w|c_k,y=l)p_\theta(c_k|y=l)
$$
where $p_\theta(w|c_k,y=l)$ captures the distribution of $w$ across $\mathcal{C}_\text{train}$ for $k^\text{th}$ concept and label $l$, while $p_\theta(c_k|y=l)$ captures the distribution of the concept $c_k$ across $\mathcal{C}_\text{train}$ for label $l$.
They can be computed using the following equations.

\begin{equation}
\aligned
p_\theta(w|c_k,y=l)&=\sum_{\xi\in\mathcal{C}_\text{train}^l}p_\theta(w,\xi|c_k,y=l),\\
p_\theta(c_k|y=l)&=\sum_{\xi\in\mathcal{C}_\text{train}^l}p_\theta(c_k,\xi|y=l),
\label{eqn:keywords_concept_corpus_level}
\endaligned
\end{equation}
where we define
\begin{equation}
p_\theta(w,\xi|c_k,y=l):=\frac{\sum_{t=1}^T\alpha^{\text{abs},\xi}_{k,t}\cdot\delta(w_t,w)}{\sum_{\xi'\in\mathcal{C}_\text{train}}f_{\xi'}(w)+\gamma}
\label{eqn:keywords_concept_case_level1}
\end{equation}
and
\begin{equation}
p_\theta(c_k,\xi|y=l):=\frac{\alpha^{\text{agg},\xi}_k}{|\mathcal{C}_\text{train}|},
\label{eqn:keywords_concept_case_level2}
\end{equation}
where $\alpha^{\text{abs},\xi}_{k,t}$ represents $\alpha^{\text{abs}}_{k,t}$ for document $\xi$.
Based on Eq.~(\ref{eqn:keywords_concept_corpus_level}), we are able to obtain scores (importance) and most relevant keywords for different concepts for a given label $l$.

\subsubsection{Consistency Analysis}
\label{sec:consistency_model}

In corpus-level and concept-based explanations, we have obtained causal relationships between keywords and predictions, i.e., $p_\theta(w|y=l)$.
However, we have not verified if these keywords are really important to predictions.
To achieve this goal, we build a Na\"ive Bayes classifier \cite{friedman1997bayesian} (NBC) based on these causal relationships.
Formally, for each testing document $d$, the probability of getting label $l$ is approximated as follows:
\begin{equation}
\aligned
p_\theta(y=l|d)&=\frac{p_\theta(d|y=l)p_\theta(y=l)}{p(d)}\\
&\propto p_\theta(d|y=l)
=\prod_{t=1}^T p_\theta(w_t|y=l),
\endaligned
\label{eqn:bayes_prediction}
\end{equation}
where $p_\theta(w_t|y=l)$ is obtained by Eq.~(\ref{eqn:keywords_corpus_level}) or Eq.~(\ref{eqn:keywords_concept_corpus_level}) on the training corpus.
We further approximate Eq.~(\ref{eqn:bayes_prediction}) with
\begin{equation}
p_\theta(y=l|d)=\prod_{w\in d'}(p_\theta(w|y=l)+\lambda),
\label{eqn:bayes_prediction_approx}
\end{equation}
where $d'\subset d$ is an \textit{attention-based bag-of-words representation} for document $d$.
It consists of important keywords based on attention weights.
$\lambda$ is a smoothing factor.
Here, we can conduct consistency analysis by comparing labels obtained by the model and NBC, which may also help estimate the uncertainty of a model \cite{zhang2019mitigating}.

%% file: experiments.tex
\section{Experiments}
\label{sec:exp}
\subsection{Datasets}

We conducted experiments on three publicly available datasets. Newsroom is used for news categorization, while IMDB and Beauty are used for sentiment analysis. The details of the three datasets are as follows:
1) \textbf{Newsroom} \cite{grusky2018newsroom}: 
The original dataset, which consists of 1.3 million news articles, was proposed for text summarization. In our experiments, we first determined the category of each article based on the URL, and then, randomly sample 10,000 articles for each of the five categories, including business, entertainment, sports, health, and technology.
2) \textbf{IMDB} \cite{maas2011learning}: 
This dataset contains 50,000 movie reviews from the IMDB website with binary (positive or negative) labels.
3) \textbf{Beauty} \cite{he2016ups}:
This dataset contains product reviews in beauty category from Amazon. We converted the original ratings (1-5) to binary (positive or negative) labels and sampled 20,000 reviews for each label. For all three datasets, we tokenized reviews using BERT tokenizer \cite{Wolf2019HuggingFacesTS} and randomly split them into train/development/test sets with a proportion of 8/1/1.
Statistics of the datasets are summarized in Table~\ref{tab:dataset_statistics}.

\begin{table}[!t]
    \centering
    \caption{Statistics of the datasets used.}
    \resizebox{0.4\linewidth}{!}{
    \begin{tabular}{|c|c|c|c|}
        \hline
        \bf Dataset & \bf \#docs & \bf Avg. Length & \bf Scale \\\hline
        Newsroom & 50,000 & 827 & 1-5\\\hline
        IMDB & 50,000 & 292 & 1-2 \\\hline
        Beauty & 40,000 & 91 & 1-2 \\\hline
    \end{tabular}}
    \label{tab:dataset_statistics}
    \vspace{-3mm}
\end{table}

\subsection{Models and Implementation Details}

We compare different classification models including several baselines, variants of our AAN model, and Na\"ive Bayes classifiers driven by a basic self-attention network \cite{serrano2019attention} (SAN) and AAN.
\begin{itemize}[leftmargin=*]
    \item \textbf{CNN} \cite{kim2014convolutional}: This model extracts key features from a review by applying convolution and max-over-time pooling operations \cite{collobert2011natural} over the shared word embedding layer.
    
    \item \textbf{LSTM-SAN}, \textbf{BERT-SAN}, \textbf{DistilBERT-SAN}, \textbf{RoBERTa-SAN}, and \textbf{Longformer-SAN}: All these models are based on the SAN framework. In LSTM-SAN, the encoder consists of a word embedding layer and a Bi-LSTM encoding layer, where embeddings are pre-loaded with 300-dimensional GloVe vectors \cite{pennington2014glove} and fixed during training.
    BERT \cite{Wolf2019HuggingFacesTS}, DistilBERT \cite{sanh2019distilbert}, RoBERTa \cite{liu2019roberta}, and Longformer \cite{beltagy2020longformer} leverage different pre-trained language models, which have 110M, 66M, 125M, 125M parameters, respectively.
    
    \item \textbf{AAN + C($c$) + Drop($r$)}: These are variants of AAN. C($c$) and Drop($r$) represent the number of concepts and dropout rate, respectively.
\end{itemize}

We implemented all deep learning models using PyTorch \cite{paszke2017automatic} and the best set of parameters are selected based on the development set.
For CNN based models, the filter sizes are chosen to be 3, 4, 5 and the number of filters is set to 100 for each size.
For LSTM based models, the dimension of hidden states is set to 300 and the number of layers is 2.
All parameters are trained with ADAM optimizer \cite{kingma2014adam} with a learning rate of 0.0002.
Dropout with a rate of 0.1 is also applied in the classification layer.
For all explanation tasks, we set the number of concepts to 10 and dropout-rate to 0.02.
Our codes and datasets are available at \url{https://github.com/tshi04/ACCE}.

\subsection{Performance Results}
\label{sec:performance}

\begin{table}[!t]
    \centering
    \caption{Averaged accuracy of different models on Newsroom, IMDB, and Beauty testing sets.}
    \resizebox{0.45\linewidth}{!}{
    \begin{tabular}{|l|c|c|c|}
        \hline
        \bf Model & \bf Newsroom & \bf IMDB & \bf Beauty \\ 
        \hline
        CNN & 90.18 & 88.56 & 88.42 \\\hline
        LSTM-SAN & 91.26 & 90.68 & 92.00 \\\hline
        BERT-SAN & 92.28 & 92.60 & 93.72 \\\hline
        DistilBERT-SAN & \bf 92.66 & 92.52 & 92.82\\\hline
        RoBERTa-SAN & 91.16 & 92.76 & 93.40\\\hline
        Longformer-SAN & 92.04 & \bf 93.74 & \bf 94.50 \\\hline
    \end{tabular}}
    \label{tab:performance-san}
\end{table}

\begin{table}[!t]
	\caption{Averaged accuracy of BERT and Longformer-based AAN models on Newsroom, IMDB, and Beauty testing sets.}
	\vspace{-2mm}
	\resizebox{0.8\linewidth}{!}{
	\begin{tabular}{|l|c|c|c|c|c|c|}
		\hline
		& \multicolumn{2}{c|}{Newsroom} 
		& \multicolumn{2}{c|}{IMDB} 
		& \multicolumn{2}{c|}{Beauty}
		\\\hline
		
		& BERT & Longformer
		& BERT & Longformer
		& BERT & Longformer
		\\\hline
		
		SAN Framework
		& 92.28 & 92.04 
		& 92.60 & 93.74 
		& 93.72 & 94.50 \\\hline
		AAN + C(10) + Drop(0.01)
		& 92.54 & 91.72 
		& 92.22 & 92.96 
		& 93.38 & 93.42
		\\\hline
		AAN + C(10) + Drop(0.02)
		& 92.14 & 91.64 
		& 92.14 & 92.86
		& 93.58 & 93.75
		\\\hline
		AAN + C(10) + Drop(0.05)
		& 92.14 & 91.60 
		& 91.82 & 92.66
		& 93.05 & 93.80
		\\\hline
		AAN + C(10) + Drop(0.10)
		& 92.30 & 91.48 
		& 91.50 & 92.12
		& 93.25 & 93.60
		\\\hline
		AAN + C(20) + Drop(0.01)
		& 92.02 & 91.98
		& 91.64 & 92.78
		& 93.70 & 93.48
		\\\hline
		AAN + C(20) + Drop(0.02)
		& 92.44 & 91.84
		& 91.80 & 93.04
		& 93.55 & 93.88
		\\\hline
		AAN + C(20) + Drop(0.05)
		& 92.54 & 91.86 
		& 91.92 & 93.14
		& 93.68 & 93.42
		\\\hline
		AAN + C(20) + Drop(0.10)
		& 92.52 & 91.98
		& 92.10 & 92.96
		& 93.72 & 93.88
		\\\hline
	\end{tabular}}
	\label{tab:performance-aan}
	\vspace{-5mm}
\end{table}

We use accuracy as evaluation metric to measure the performance of different models.
All quantitative results have been summarized in Tables~\ref{tab:performance-san} and \ref{tab:performance-aan}, where we use bold font to highlight the highest accuracy on testing sets in Table~\ref{tab:performance-san}.
Comparing LSTM-SAN with BERT, DistilBERT, RoBERTa and Longformer, we first find that different pre-trained language model-based encoders are better than conventional LSTM encoder with pre-trained word embeddings.
In Table~\ref{tab:performance-aan}, We replace self-attention on top of pre-trained language models with abstraction-aggregation network (AAN).
We observe that different AAN models do not significantly lower the classification accuracy, which indicates we can use AAN for the concept-based explanation task without losing the overall performance.
Here, the strategy of aggregation-attention weights dropout is necessary when training AAN models.
In Table~\ref{tab:dropout_example}, we show that
AAN models without randomly dropping aggregation-attention weights attain poor interpretability in concept-based explanation.

\subsection{Heat-maps and Case-level Concept-based Explanation}
\label{sec:exp_case_level_explanation}

\begin{table}[!t]
    \centering
    \caption{Case-level concept-based explanation. Here, each ID is associated with a concept, i.e., abstraction-attention unit. Scores and weights (following each keyword) are calculated with Eq.~(\ref{eqn:keywords_concept_case_level1}) and (\ref{eqn:keywords_concept_case_level2}). `-' represents special characters.}
    \resizebox{0.7\linewidth}{!}{
    \begin{tabular}{|c|c|m{24em}|}
    \hline
    \bf ID & \bf Score & \bf Keywords\\
    \hline
    8 & 0.180 & com(0.27), boston(0.26), boston(0.16), boston(0.1), m(0.02) \\\hline
    6 & 0.162 & marketing(0.28), ad(0.06), \#\#fs(0.05), investors(0.03), said(0.03) \\\hline
    1 & 0.148 & campaign(0.14), firm(0.14), money(0.06), brand(0.04), economist(0.03) \\\hline
    2 & 0.122 & economist(0.2), said(0.16), professional(0.16), investors(0.08), agency(0.06) \\\hline
    9 & 0.116 & boston(0.89), boston(0.11) \\\hline
    7 & 0.108 & bloomberg(0.16), cn(0.11), global(0.09), money(0.06), cable(0.05) \\\hline
    5 & 0.103 & -(0.96), -(0.03), s(0.01) \\\hline
    4 & 0.047 & investment(0.76), money(0.14), investment(0.06), investment(0.02), investors(0.01) \\\hline
    3 & 0.016 & ,(0.64), -(0.36) \\\hline
    10 & 0.000 & .(0.93), -(0.07) \\\hline
    \end{tabular}}
    \label{tab:case_level_concept_explanation}
    \vspace{-3mm}
\end{table}

\begin{figure}[!t]
    \centering
    \includegraphics[width=0.6\linewidth]{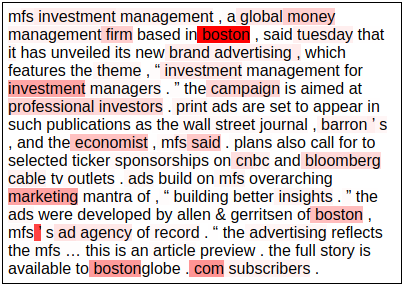}
    \caption{Attention-weight visualization for an interpretable attention-based classification model.}
    \label{fig:introduction_example}
    \vspace{-5mm}
\end{figure}

First, we investigate if AAN attends to relevant keywords when it is making predictions, which can be accomplished by visualizing attention weights (see Fig.~\ref{fig:introduction_example}).
This is a \textit{Business} news article from Newsroom and we observe that the most relevant keyword that AAN detects is \textit{boston}.
Other important keywords include \textit{investment}, \textit{economist}, \textit{marketing} and \textit{com}.
Compared with Fig.~\ref{fig:introduction_example}, our \textit{case-level concept-based explanation} provides more informative results.
From Table~\ref{tab:case_level_concept_explanation}, we observe that AAN makes the prediction based on several different aspects, such as corporations (e.g., com), occupations (e.g., economist), terminology (e.g., marketing) and so on.
Moreover, \textit{boston} may be related with corporation (e.g., bostonglobe or gerritsen of boston) or city, thus, it appears in both concepts 8 (corporations) and 9 (locations).

\subsection{Corpus-Level Explanation}

\begin{figure}[!t]
    \centering
    \begin{subfigure}{\linewidth}
    \centering
    \includegraphics[width=0.4\linewidth]{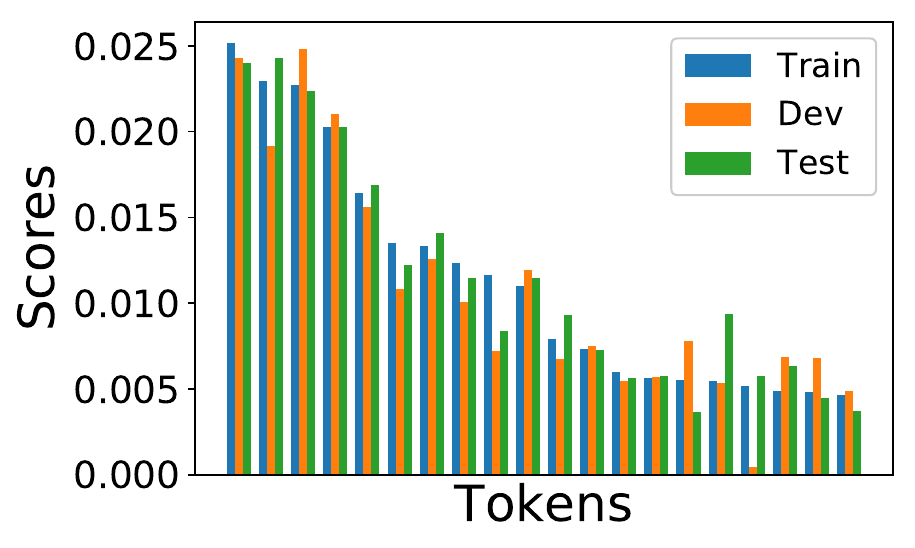}
    \includegraphics[width=0.4\linewidth]{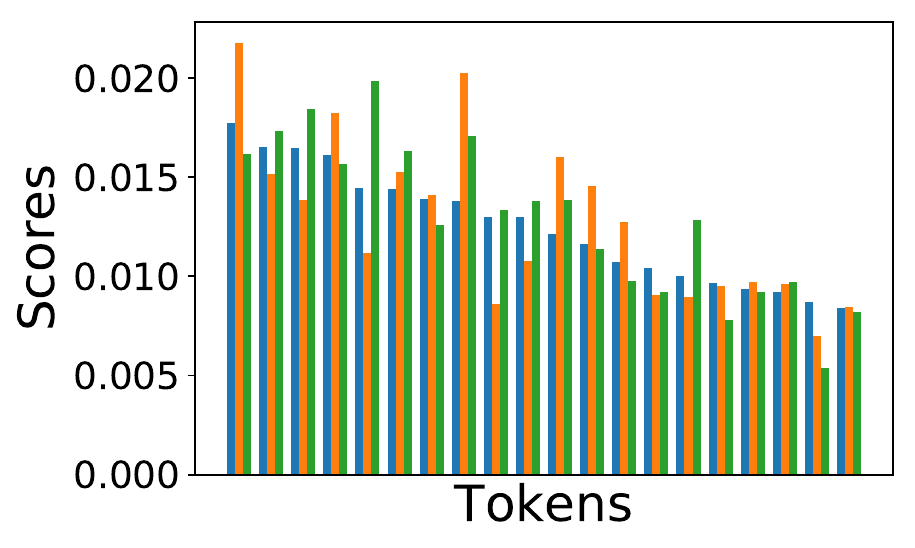}
    \caption{Newsroom. Left: Business. Right: Sports}
    \end{subfigure}
    \begin{subfigure}{\linewidth}
    \centering
    \includegraphics[width=0.4\linewidth]{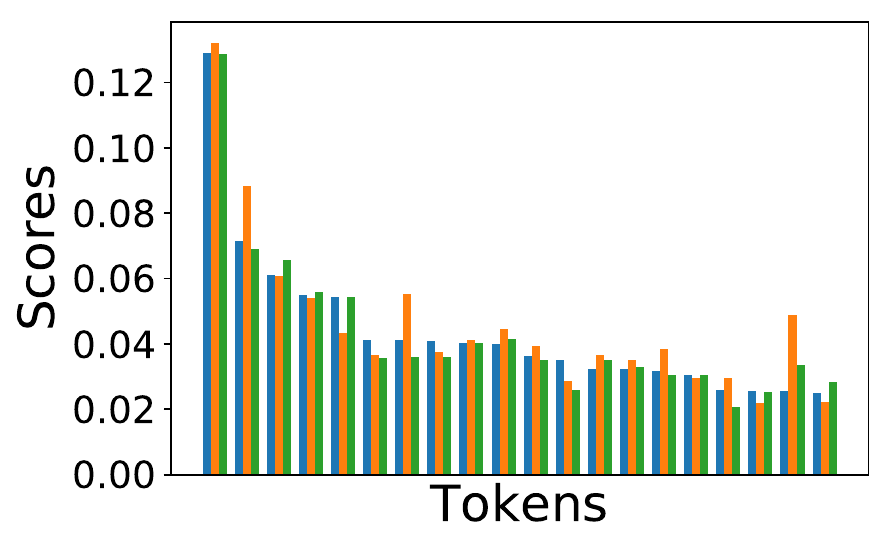}
    \includegraphics[width=0.4\linewidth]{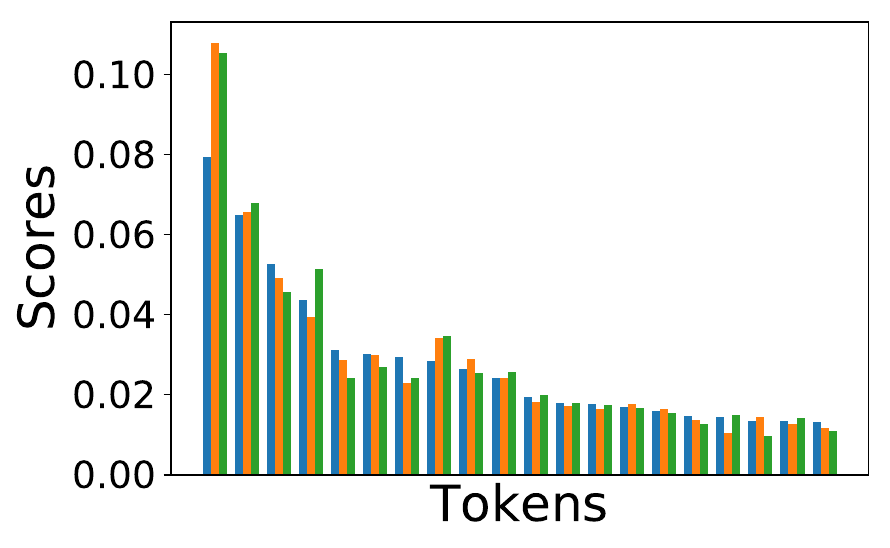}
    \caption{IMDB. Left: Negative. Right: Positive}
    \end{subfigure}
    \begin{subfigure}{\linewidth}
    \centering
    \includegraphics[width=0.4\linewidth]{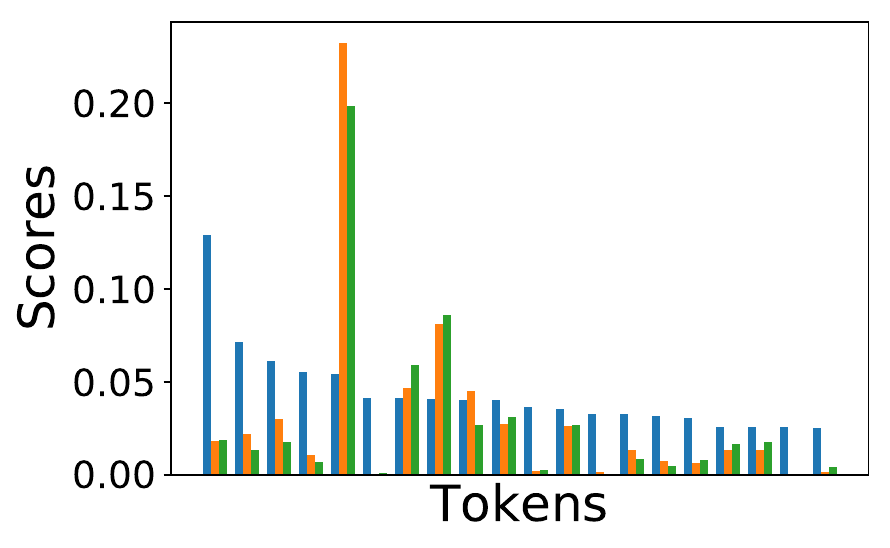}
    \includegraphics[width=0.4\linewidth]{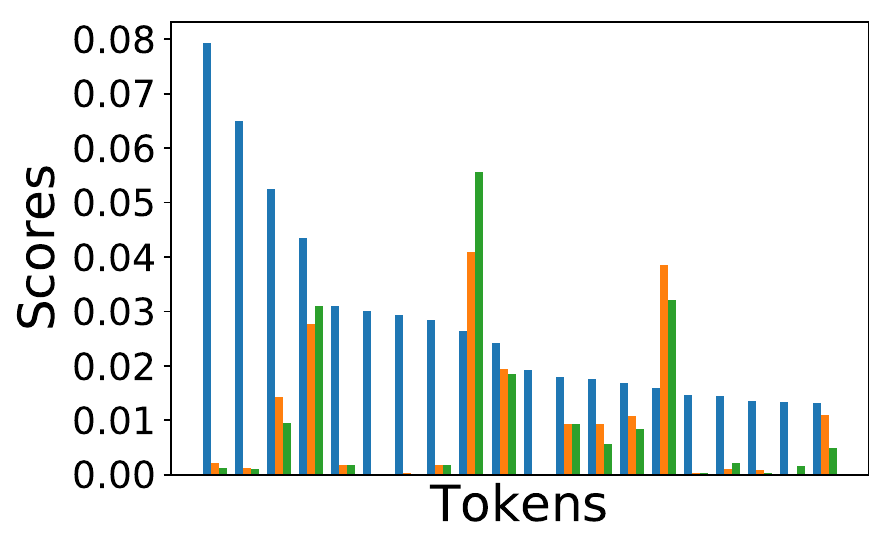}
    \caption{IMDB-to-Beauty. Left: Negative. Right: Positive}
    \end{subfigure}
    \caption{Distribution of keywords on training, development and testing sets. Scores are calculated by $p_\theta(w,l,\mathcal{C})$. The orders of tokens are the same as those in Table~\ref{tab:keywords_corpus_level}.}
    \label{fig:distribution_keywords}
    \vspace{-3mm}
\end{figure}

Corpus-level explanation aims to find the important keywords for the predictions.
In Table~\ref{tab:keywords_corpus_level}, we show 20 most important keywords for each predicted label and we assume these keywords determine the predictions.
In the last section, we will demonstrate this assumption by the consistency analysis.
The scores of keywords have been shown in Fig.~\ref{fig:distribution_keywords}.

In addition to casual relationships, we can also use these keywords to check if our model and datasets have bias or not.
For example, \textit{boston} and \textit{massachusetts} plays an important role in predicting business, which indicates the training set has bias.
By checking our data, we find that many business news articles are from \textit{The Boston Globe}.
Another obvious bias example is that the numbers \textit{8, 7}  and \textit{9} are important keywords for IMDB sentiment analysis.
This is because the original ratings scale from 1 to 10 and many reviews mention that ``\textit{rate this movie 8 out of 10}''.

\begin{table}[!t]
	\centering
	\caption{This table shows 20 most important keywords for model predictions on different training sets. Keywords are ordered by their scores. For Newsroom, we only show 2 out of 5 classes due to space limitations.}
	\resizebox{\linewidth}{!}{
	\begin{tabular}{|c|c|m{32em}|}
    \hline
    \bf Dataset & \bf Label & \bf Keywords
    \\\hline
    \multirow{4}{*}{IMDB} 
    & Negative & 
    worst, awful, terrible, bad, \textbf{disappointed}, boring, \textbf{disappointing}, \textbf{waste}, \textbf{horrible}, sucks, fails, disappointment, lame, dull, poorly, poor, worse, mess, dreadful, pointless
    \\\cline{2-3}
    & Positive & 
    8, 7, \textbf{excellent}, \textbf{loved}, 9, enjoyable, superb, enjoyed, \textbf{highly}, \textbf{wonderful}, entertaining, best, beautifully, good, \textbf{great}, brilliant, terrific, funny, hilarious, fine
    \\\hline
    
    \multirow{4}{*}{Beauty} 
    & Negative &
    disappointed, nothing, unfortunately, made, not, waste, disappointing, terrible, worst, horrible, makes, no, sadly, disappointment, t, awful, sad, bad, never, started
    \\\cline{2-3}
    & Positive & 
    great, love, highly, amazing, pleased, perfect, works, best, happy, awesome, makes, recommend, excellent, wonderful, definitely, good, glad, well, fantastic, very
    \\\hline
    
    \multirow{13}{*}{Newsroom}
    & Business &
    inc, corp, boston, massachusetts, economic, cambridge, financial, economy, banking, auto, automotive, startup, company, mr, finance, biotechnology, somerville, retailer, business, airline
    \\\cline{2-3}
    & Entertainment & 
    singer, actress, actor, star, fox, comedian, hollywood, sunday, rapper, fashion, celebrity, contestant, filmmaker, bachelor, insider, porn, oscar, rocker, host, monday
    \\\cline{2-3}
    & Sports & 
    quarterback, coach, basketball, baseball, soccer, nba, sports, striker, tennis, hockey, nfl, nhl, football, olympic, midfielder, golf, player, manager, outfielder, nascar
    \\\cline{2-3}
    & Health & 
    dr, health, pediatric, obesity, cardiovascular, scientists, researcher, medicine, psychologist, diabetes, medical, psychiatry, aids, fitness, healthcare, autism, psychology, neuroscience, fox, tobacco
    \\\cline{2-3}
    & Technology & 
    tech, cyber, electronics, wireless, lifestyle, silicon, gaming, culture, telecommunications, scientist, company, google, smartphone, technology, francisco, broadband, privacy, internet, twitter
    \\\hline
	\end{tabular}}
	\label{tab:keywords_corpus_level}
	\vspace{-5mm}
\end{table}

Moreover, from Fig.~\ref{fig:distribution_keywords} (a) and (b), we find that for a randomly split corpus, distributions of keywords across training/development/test sets are similar to each other.
This guarantees the model achieve outstanding performance on testing sets.
If we apply a model trained on IMDB to Beauty (see Fig.~\ref{fig:distribution_keywords} (c)), it can only leverage the cross-domain common keywords (e.g., \textit{disappointed} and \textit{loved}) to make predictions.
However, we achieve $71\%$ accuracy, which is much better than random predictions.
In Table~\ref{tab:keywords_corpus_level}, we use bold font to highlight these common keywords.

\subsection{Corpus-level Concept-based Explanation}
\label{sec:concept_explanation}

Corpus-level concept-based explanation further improves the corpus-level explanation by introducing clustering structures to keywords.
In this section, we still use AAN trained on Newsroom as an example for this task.
Table~\ref{tab:corpus_level_explanation_concept_business} shows concepts and relevant keywords for AAN when it assigns an article to \textit{Business}.
Here, we observe that the first-tier salient concepts consist of concepts 8 (corporations) and 1 (business terminology in general).
The second-tier concepts 7, 6, and 4 are related to economy, finance, mortgage, and banking, which are domain-specific terminology.
They share many keywords.
Concepts 9 and 2 are associated with locations and occupations, respectively, which receive relatively lower scores.
Concepts 5, 3 and 10 are not quite meaningful.
We have also shown results for Newsroom sports in Table~\ref{tab:corpus_level_explanation_concept_sports}, where we find that 1 (sports terminology) and 7 (leagues and teams) are the first-tier salient concepts.
The second-tier salient concepts 6 and 4 are about games and campaigns.
Concepts 7, 6, 4 also share many keywords.
Concepts 8 (corporations and channels), 2 (occupations and roles) and 9 (locations) are the third-tier salient concepts.
Concepts 5, 3, 10 are also meaningless.
From these tables, we summarize some commonalities: 
1) Domain-specific terminologies (i.e., concepts 1, 7, 6 and 4) play an important role in predictions.
2) Locations (i.e., concept 9) and Occupations/Roles (i.e., concept 2) are less important. 
3) Meaningless concepts (i.e., concepts 5, 3, and 10), such as punctuation, have the least influence.

\begin{table}[!t]
    \centering
    \caption{Concept-based explanation (Business). Scores are calculated using Eq.~(\ref{eqn:keywords_concept_corpus_level}).}
    \resizebox{\linewidth}{!}{
    \begin{tabular}{|c|c|m{40em}|}
    \hline
    \bf ID & \bf Score & \bf Keywords\\
    \hline
    8 & 0.173 & inc, corp, massachusetts, boston, mr, ms, jr, ltd, mit, q \\\hline
    1 & 0.168 & economy, retailer, company, startup, \#\#maker, airline, chain, bank, utility, billionaire \\\hline
    7 & 0.151 & biotechnology, banking, tech, startup, pharmaceuticals, mortgage, financial, auto, commerce, economic \\\hline
    6 & 0.124 & economic, health, banking, finance, insurance, healthcare, economy, housing, safety, commerce \\\hline
    4 & 0.107 & financial, economic, banking, auto, automotive, securities, housing, finance, monetary, biotechnology \\\hline
    9 & 0.086 & boston, massachusetts, cambridge, washington, detroit, frankfurt, harvard, tokyo, providence, paris \\\hline
    5 & 0.056 & -, \#\#as, -, -, itunes, inc, corp, northeast, -, llc \\\hline
    2 & 0.054 & economist, executive, spokesman, analyst, economists, \#\#gist, ceo, director, analysts, president \\\hline
    3 & 0.026 & -, -, -, ), \#\#tem, \#\#sp, the, =, t, ob \\\hline
    10 & 0.000 & -, comment, ), insurance, search, ', tesla, graphic, guitarist, , \\\hline
    \end{tabular}}
    \label{tab:corpus_level_explanation_concept_business}
\end{table}

\begin{table}[!t]
    \centering
    \caption{Concept-based explanation (Sports). Scores are calculated using Eq.~(\ref{eqn:keywords_concept_corpus_level}).}
    \resizebox{\linewidth}{!}{
    \begin{tabular}{|c|c|m{40em}|}
    \hline
    \bf ID & \bf Score & \bf Keywords\\
    \hline
    1 & 0.176 & quarterback, player, striker, champion, pitcher, midfielder, outfielder, athlete, goaltender, forward \\\hline
    7 & 0.165 & nhl, mets, soccer, nets, yankees, nascar, mls, reuters, doping, twitter \\\hline
    6 & 0.147 & tennis, sports, soccer, golf, doping, hockey, athletic, athletics, injuries, basketball \\\hline
    4 & 0.139 & baseball, basketball, nba, nfl, sports, football, tennis, olympic, hockey, golf \\\hline
    8 & 0.119 & jr, ", n, j, fox, espn, nl, u, boston, ca \\\hline
    2 & 0.100 & coach, manager, commissioner, boss, gm, trainer, spokesman, umpire, coordinator, referee \\\hline
    9 & 0.060 & philadelphia, indianapolis, boston, tampa, louisville, buffalo, melbourne, manchester, baltimore, atlanta \\\hline
    5 & 0.055 & ’, ', \#\#as, –, \#\#a, `, sides, newcomers, chelsea, jaguars \\\hline
    3 & 0.022 & ’, ', ), ,, \#\#kus, \#\#gre, the, whole, lever, \#\#wa \\\hline
    10 & 0.000 & ., ), finishes, bel, gymnastics, ', \#\#ditional, becomes, tu, united \\\hline
        \end{tabular}}
    \label{tab:corpus_level_explanation_concept_sports}
\end{table}

\subsection{Consistency Analysis}
\label{sec:consistent}

In this section, we leverage the method proposed in Section~\ref{sec:consistency_model} to respectively build a NBC for BERT-SAN and BERT-AAN on the training set.
Then, we apply them to the testing set to compare if NBC predictions and the model predictions are consistent with each other.
We approximate the numerator of Eq.~(\ref{eqn:keywords_case_level}) with five words (can repeat) with highest attention weights in each document.
In Eq.~(\ref{eqn:keywords_corpus_level}), $\gamma$ is set to be 1000.
In Eq.~(\ref{eqn:bayes_prediction_approx}), we set $\lambda=1.2$ for text categorization and $\lambda=1.0$ for sentiment analysis.
$d'$ consists of five words with highest attention weights.

We use the accuracy (consistency score) between labels predicted by NBC and the original model to evaluate the consistency.
Table~\ref{tab:consistency} shows that around $85\%$ of predictions are consistent.
This demonstrates that keywords obtained by the corpus-level and concept-based explanation methods are important to predictions.
They can be used to interpret attention based models.
Moreover, from CP and NCP scores, we observe a significantly higher probability that the model makes an incorrect prediction if it is inconsistent with NBC prediction.
This finding suggests us to use consistency score as one criterion for \textit{uncertainty estimation}.

\subsection{Dropout of Aggregation-Attention Weights}

For AAN, we apply dropout to aggregation-attention weights during training.
In Table~\ref{tab:dropout_example}, we show an example without using the attention weight dropout mechanism.
We observed that the weight for concept 1 is much higher than other concepts.
In addition, keywords for each concept are not semantically coherence.

\begin{table}[!t]
    \centering
    \caption{Consistency between the model and NBC. CS represents consistency score, CP/NCP denote percentage of incorrect predictions when NBC predictions are consistent/not consistent with model predictions.}
    \vspace{-3mm}
    \resizebox{0.7\linewidth}{!}{
    \begin{tabular}{|l|c|c|c|c|c|c|c|c|c|}
        \hline
        \bf Model & 
        \multicolumn{3}{c|}{\bf Newsroom} & 
        \multicolumn{3}{c|}{\bf IMDB} & 
        \multicolumn{3}{c}{\bf Beauty} \\ 
        \hline
        &CS&NCP&CP&CS&NCP&CP&CS&NCP&CP\\\hline
        BERT-SAN & 83.96 & 21.59 & 4.72 & 86.02 & 17.17 & 5.81 & 85.45 & 16.30 & 4.56\\\hline
        BERT-AAN & 84.36 & 20.20 & 5.57 & 85.46 & 21.18 & 5.05 & 84.72 & 16.04 & 4.51\\
        \hline
    \end{tabular}}
    \label{tab:consistency}
    \vspace{-3mm}
\end{table}

\begin{table}[!t]
    \centering
    \caption{Concept-based explanation (Sports) for AAN without applying dropout to attention weights.}
    \resizebox{1.\linewidth}{!}{
    \begin{tabular}{|c|c|m{35em}|}
    \hline
    \bf CID & \bf Weight & \bf Keywords\\
    \hline
    1 & 0.8195 & quarterback, athletic, olympic, basketball, athletics, qb, hockey, outfielder, sports \\\hline
    7 & 0.0865 & nascar, celtics, motorsports, nba, boston, augusta, nhl, tennis, leafs, zurich \\\hline
    4 & 0.0370 & mets, knicks, yankees, players, pitchers, lakers, hosts, coaches, forwards, swimmers \\\hline
    3 & 0.0164 & offensive, eli, bird, doping, nba, jay, rod, hurdle, afc, peyton \\\hline
    2 & 0.0098 & premier, american, mets, nl, field, yankee, national, aaron, nba, olympic \\\hline
    10 & 0.0083 & games, seasons, tries, defeats, baskets, players, season, contests, points, throws \\\hline
    5 & 0.0015 & dustin, antonio, rookie, dante, dale, dylan, lineman, ty, launch, luther \\\hline
    8 & 0.0010 & 2016, 2014, college, tribune, card, press, s, -, this, leadership \\\hline
    9 & 0.0004 & men, -, grand, 9, s, usa, state, west, world, major \\\hline
    6 & 0.0000 & the, -, -, year, whole, vie, very, tr, too, to \\\hline
    \end{tabular}}
    \label{tab:dropout_example}
\end{table}

%% file: conclusion.tex
\section{Conclusion}
\label{sec:conclusion}

In this paper, we proposed a general-purpose \textit{corpus-level explanation} approach to interpret attention-based networks.
It can capture causal relationships between keywords and model predictions via learning importance of keywords for predicted labels across the training corpus based on attention weights.
Experimental results show that the keywords are semantically meaningful for predicted labels.
We further propose a \textit{concept-based explanation} method to identify important concepts for model predictions.
This method is based on a novel \textit{Abstraction-Aggregation Network} (AAN), which can automatically extract concepts, i.e., clusters of keywords, during the end-to-end training for the main task.
Our experimental results also demonstrate that this method effectively captures semantically meaningful concepts/clusters.
It also provides relative importance of each concept to model predictions.
To verify our results, we also built a \textit{Na\"ive Bayes Classifier} based on an \textit{attention-based bag-of-word document representation} technique and the casual relationships.
Consistency analysis results demonstrate that the discovered keywords are important to the predictions.